\documentclass[conference]{IEEEtran}
\usepackage{cite}

\ifCLASSINFOpdf
  \usepackage[pdftex]{graphicx}
\else
  \usepackage[dvips]{graphicx}
\fi
\usepackage{amsmath}
\ifCLASSOPTIONcompsoc
 \usepackage[caption=false,font=normalsize,labelfont=sf,textfont=sf]{subfig}
\else
 \usepackage[caption=false,font=footnotesize]{subfig}
\fi
\usepackage[binary-units=true]{siunitx}

\usepackage{hyphenat}
\usepackage[acronym]{glossaries}
\newacronym{cim}{CIM}{Computation\hyp{}in\hyp{}Memory}
\newacronym{nfa}{NFA}{Non\hyp{}deterministic Finite Automata}
\newacronym{ste}{STE}{State Transition Element}
\newacronym{mvp}{MVP}{Memristive Vector Processor}

\usepackage[final]{changes}

\begin{document}
%
\title{Memristive Devices for Computation-In-Memory}

\author{\IEEEauthorblockN{\makebox[\linewidth][c]}{Jintao Yu, Hoang Anh Du Nguyen, Lei Xie, Mottaqiallah Taouil, Said Hamdioui}
\IEEEauthorblockA{Laboratory of Computer Engineering, Delft University of Technology, the Netherlands \\
Email: \{J.Yu-1,H.A.DuNguyen,L.Xie,M.Taouil,S.Hamdioui\}@tudelft.nl} }

\maketitle

\begin{abstract}
CMOS technology and its continuous scaling have made electronics and computers accessible and affordable for almost everyone on the globe; in addition, they have enabled the solutions of a wide range of societal problems and applications.  
 Today, however, both the technology and the
computer architectures are facing severe challenges/walls making them incapable of providing the demanded computing
power with tight constraints. This motivates the need for
the exploration of novel architectures based on new device technologies; not only to
sustain the financial benefit of technology scaling, but also to develop 
solutions for extremely demanding emerging applications. 
This paper presents two computation-in-memory based accelerators making use of emerging memristive devices; they are Memristive Vector Processor and RRAM Automata Processor. 
The preliminary results of these two accelerators show significant improvement in terms of latency, energy  and area as compared to today's architectures and design.
\end{abstract}


%
\IEEEpeerreviewmaketitle

\section{Introduction}
Today's and new emerging applications, such as data-intensive/big-data applications (e.g., DNA sequencing) and internet-of-things (IoT), are extremely demanding with respect to computing power, energy consumption, and storage. These applications will not only strongly shape our near future, but also impact the semiconductor and computer industry. However, their requirements are difficult to fulfill with today's CMOS based computer architectures, as they face sever challenges both at architectural and device level. Current computer architectures face three walls \cite{hennessy2011computer}: (1) the memory wall due to the growing gap between processor and memory speed and the the limited memory bandwidth; (2) the power wall as the practical power budget for cooling has been reached; (3) the instruction-level parallelism (ILP) wall due to the growing difficulties in extracting enough parallelism in software/code that can run on the mainstream parallel hardware today. The CMOS devices also face three walls \cite{hamdioui2017memristor}: (1) the leakage wall as the static power is becoming dominant at small technology nodes (due to volatile technology and low Vdd) and it may even be higher than the dynamic power, (2) the reliability wall as technology scaling leads to reduced device lifetime and higher failure rate; (3) the cost wall as the cost per device from a pure geometric scaling of technology point of view is plateauing. Both architecture and device walls have slowed down the performance gains of CMOS-based architectures. All these motivate the need to look for alternative architectures while considering emerging device technologies. 
 
Many alternatives architectures are under investigations.  Resistive computing~\cite{hamdioui2015memristor, Gaillardon2016, Li2016} and neuromorphic computing architectures~\cite{yang2013memristive,furber2016large} using memristive devices, and quantum computing using quantum dots~\cite{fu2016heterogeneous} are couple of examples.  Resistive computing architectures based on memristive devices are attractive, as they enable in-memory computing (reducing the memory wall)~\cite{hamdioui2017memristor,du2017implementation}. In addition, the memristive devices have zero standby power~\cite{yang2013memristive} (helps reducing both the leakage and power wall), great scalability (reduces the cost wall),  high density (reduces the cost wall), and they are CMOS compatible (reduces the cost wall). 

This paper discusses two memristive device based accelerators to demonstrate how computation-in-memory architectures can realize significant improvements, due both to the architecture itself as well as to the used technology to implement them. First, a memristive based vector processor, referred to as \gls{mvp}, is presented; MVP can be used as an accelerator for conventional machines and shows approximately one order of magnitude improvement in performance and energy efficiency. Thereafter, a general model for hardware-based automata processing is introduced and implemented with memristive devices. This implementation is referred to as RRAM-AP; RRAM-AP's key kernel (i.e., the vector dot product operator) outperforms the state-of-the-art SRAM-based implementation by 40\% less delay and 27\% less energy, at even smaller chip area.
 

The reminder of this paper is organized as follows. Section~\ref{sec:basics} describes briefly the fundamentals of memristive devices. Section~\ref{sec:MVP} and~\ref{sec:MAP} present MVP and RRAM-AP, respectively. Finally, Section~\ref{sec:conclusion} concludes the paper.

\section{Basics of Memristive Devices} \label{sec:basics}
\begin{figure}
    \centering
    \includegraphics[width=\columnwidth]{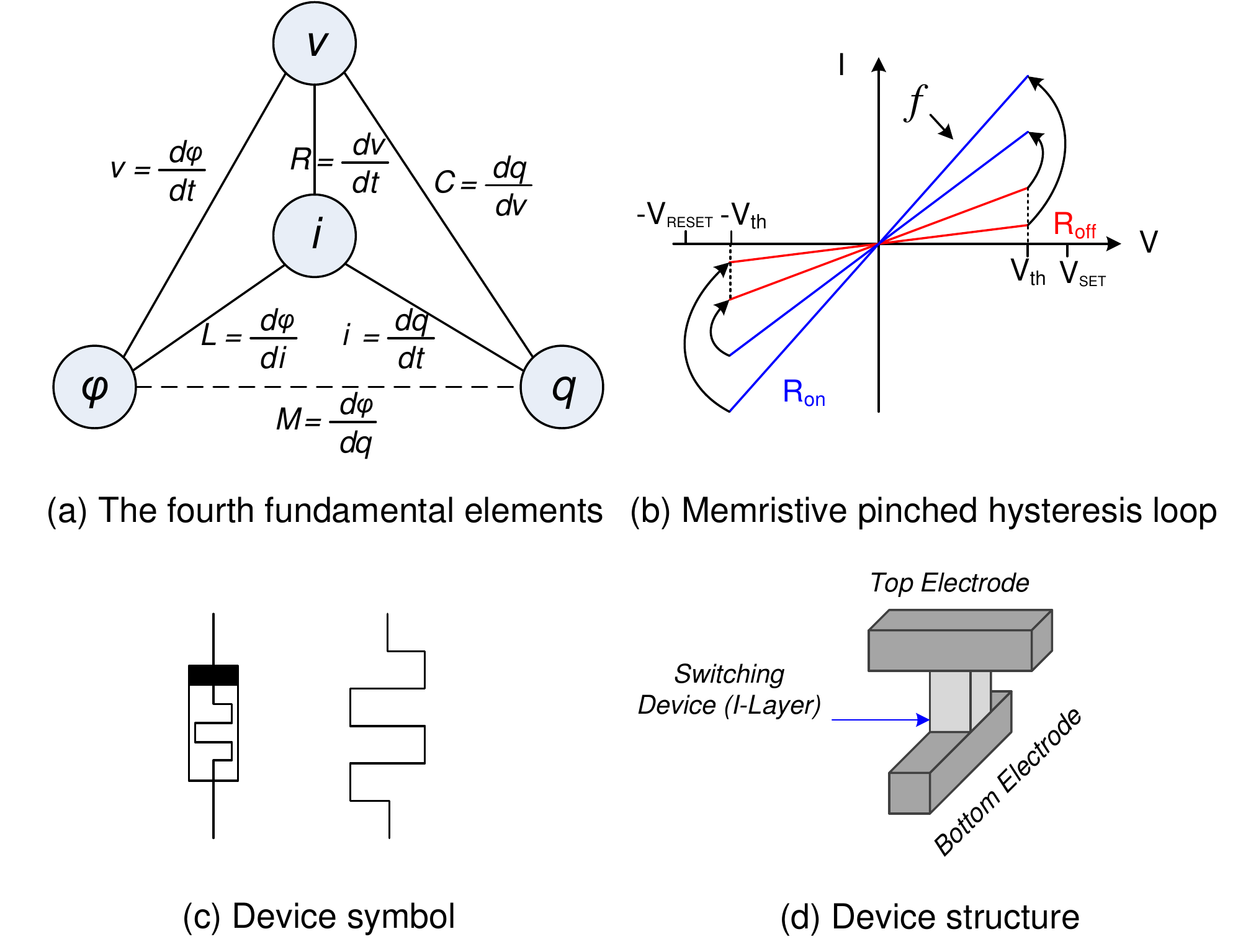}
       \caption{Main characteristics of a memristive device.}
    \label{fig:basics}
\end{figure}

The memristive device, or \textit{memristor} for short, is the fourth type of fundamental two-terminal electrical components, next to the resistor, capacitor, and inductor. It was initially predicted in 1971 by the circuit theorist Leon Chua~\cite{chua1971memristor}. 
He observed a missing element that can be described as a function of flux $\phi$ and charge $q$, as shown (with the dashed line) in Fig.~\ref{fig:basics}a. In theory, a memristive device is a passive element that can be described by the current integral (charge $q$) through or voltage integral (flux $\phi$) across its two terminals; The beauty of the
memristive device is its ability to memorize the history (i.e., the
internal state). The essential fingerprint of memristive devices is the
‘pinched current-voltage hysteresis loop’, as illustrated in Fig.~\ref{fig:basics}b. When a memristive device is floating or when the voltage \textit{v(t)} across it equals zero, the current \textit{i(t)} is also zero. Therefore, based on its hysteresis curve, the memristor has at least two distinctive states: a high ($R_H$) and low ($R_L$) resistive state. A memristive device switches from high (low) to low (high) state by applying a voltage $V_\text{SET}$ ($V_\text{RESET}$) with an absolute value larger than its threshold voltage $V_\text{th}$. Another signature of the memristive devices is that the ‘pinched hysteresis loop’ shrinks with a higher
excitation frequency f as shown in Fig.~\ref{fig:basics}b. Fig.~\ref{fig:basics}c shows the two typical symbols used to denote memristive devices; the black square represents the positive terminal.

After a silent period for more than thirty years, a practical memristive device was fabricated and demonstrated by HP in 2008~\cite{strukov2008missing}. HP built a metal-insulator-metal device using titanium oxide as an insulator and identified the memristive behaviour over its two-terminal node as described by Leon Chua; as shown in Fig.~\ref{fig:basics}d. The device resistance is modulated by controlling
positive charged oxygen vacancies in the insulator layer using different voltages. After the first memristive device was fabricated, several memristor devices based on different types of materials have been proposed such as spintronic, amorphous silicon, and ferroelectric memristors~\cite{yang2013memristive}.


\section{Memristive Devices for Vector Processing}
\label{sec:MVP}

Memristor-based Computation-In-Memory (CIM) concept was proposed to eliminate the communication between the CPU and memory by leveraging memristors for both storage and computation in the same physical crossbar~\cite{hamdioui2015memristor,barbareschi2017memristive, dunguyen2017memristive}. Here, we use the CIM to realize an accelerator we refer to as  Memristive Vector Processor (MVP). The rest of this section will describe the working principle of \gls{mvp}, the targeted applications and some analytical evaluation results to show the potential of such an architecture.

\begin{figure}
    \centering
    \includegraphics[width=0.8\columnwidth]{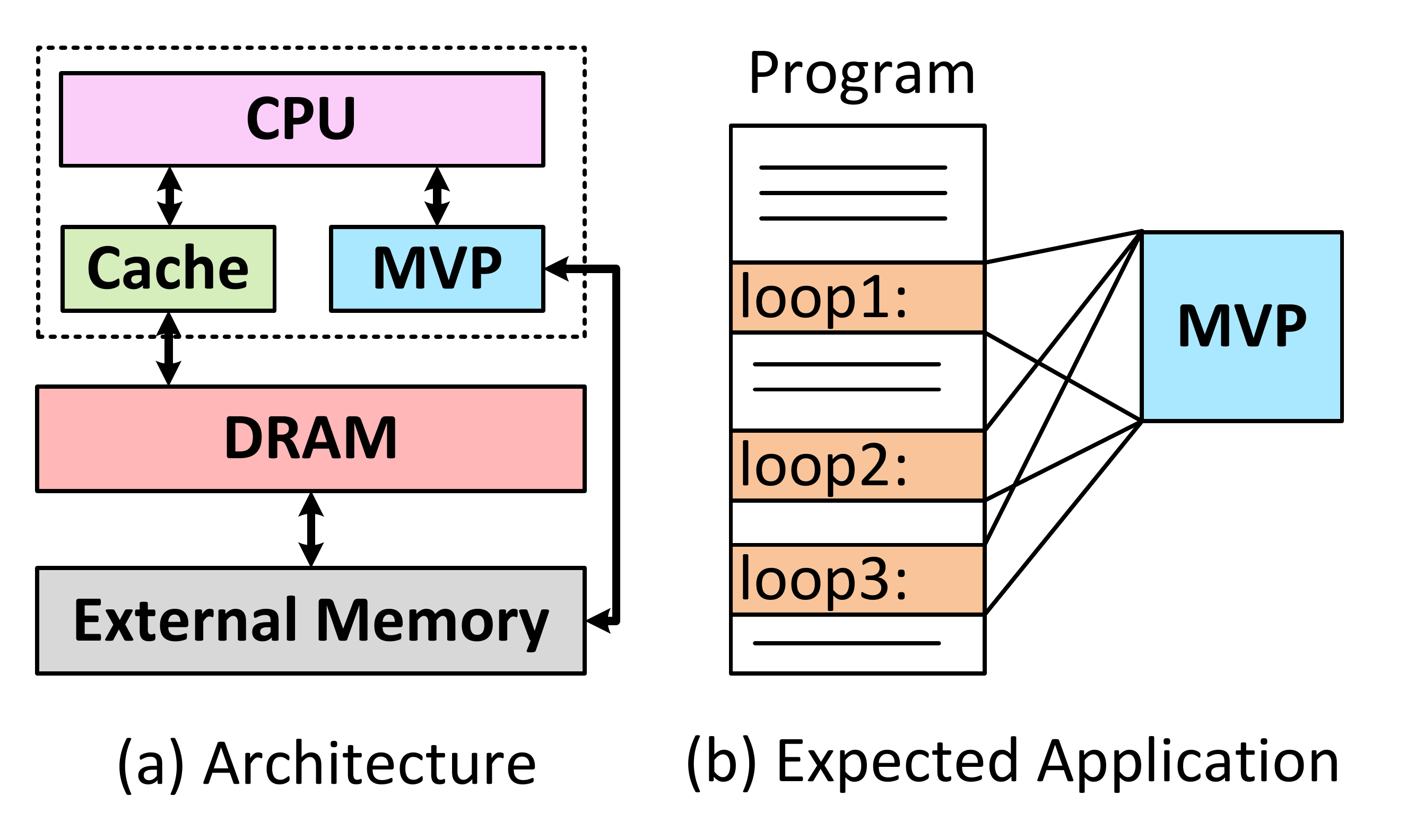}
       \caption{Memristive Vector Processor architecture.}
    \label{fig:cima}
\end{figure}

\begin{figure}[!t]
\centering
\includegraphics[width=\columnwidth]{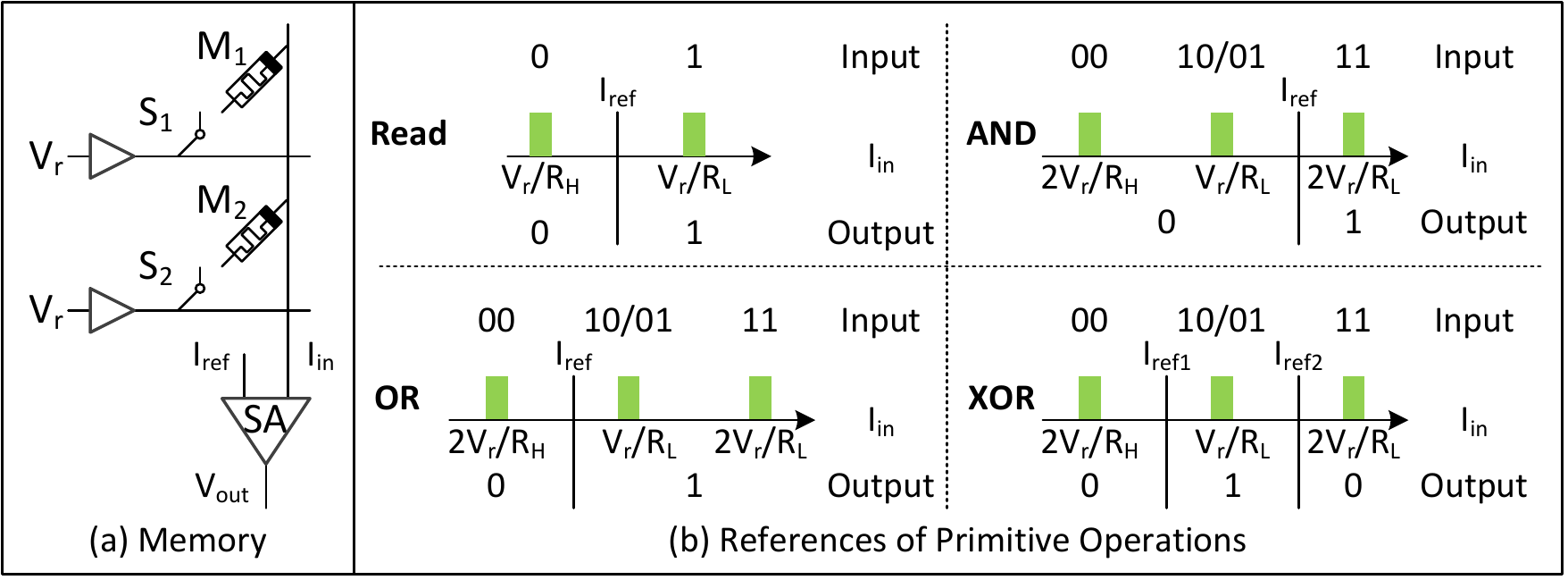}
\caption{Scouting logic \cite{xie2017scouting}.}
\label{fig:idea}
\end{figure}

\begin{figure*}
    \centering 
    \includegraphics[width=\textwidth]{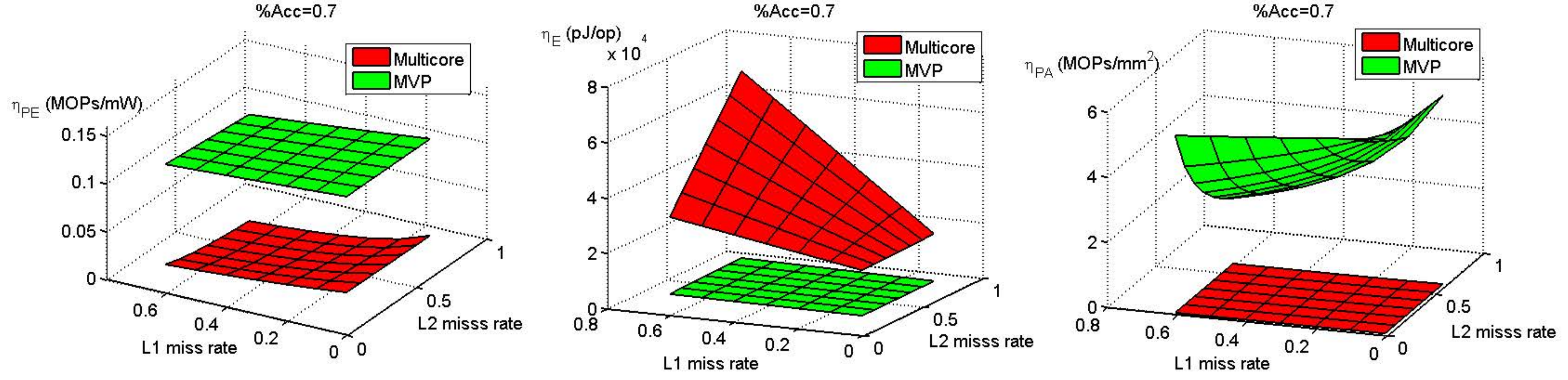}
       \caption{Evaluation results for \gls{mvp} and multicore architectures.}
    \label{fig:hpc}
\end{figure*}

\subsection{Working principle}

\gls{mvp} is proposed to accelerate applications with a huge number of vector operations. It can be used as an accelerator for a conventional processor, as shown in Fig. \ref{fig:cima}a. Similarly as in conventional architectures, the processor fetches, decodes and executes a program using a memory hierarchy consisting of cache(s), DRAM, and external memory. The part of the program which is memory intensive will be offloaded to \gls{mvp}. The distinct feature of \gls{mvp} is its crossbar memory implementation using memristive devices, which enables not the storage of huge amount of data (due to its nano scale size), but also the processing of operations within the memory (i.e., no need for data movement). 

The processing in \gls{mvp} is performed based on scouting logic operations~\cite{xie2017scouting,Li2016} ; they transform memory read operations into  logical operations. Normally, when a memory cell is being read, a read voltage $V_r$ is applied to the activated row as shown in Fig. \ref{fig:idea}a. Subsequently, a current will flow through the bit line to the input of the sense amplifier (SA) where it is compared to a reference current. Depending on the cell value (either low ($R_L$) or high  ($R_H$) resistance), the output of the SA will produce either logic 1 or 0. Inspired by this read operation, scouting logic is able to implement OR, AND and XOR gates. Instead of reading a single memristor at a time, scouting logic activates two (or more) memory rows simultaneously. As a result, the input current to the sense amplifiers is determined by the equivalent input resistance of the activated rows. This resistance results in three possible values: $R_H$, $R_H$//$R_L$ $\approx$ $R_L$, or $R_L/2$; by changing the reference current of the SA, different gates can be realized (as shown in Fig. \ref{fig:idea}b). Therefore, using this scheme allows \gls{mvp} to perform logical operations by just a small modification of the peripheral circuit of the crossbar mememory.  It eliminates the necessity of temporary registers, loading latency and energy to move data from memory to registers. It also increases the parallelism of the architecture and does not impact the the endurance of the memristive devices.

\subsection{Potential targeted applications}

With its unique capability, \gls{mvp} is able to accelerate data intensive applications. These applications consist of intensive memory accesses that consume an enormous amount of energy and degrade the overall performance due to data movements through the memory hierarchy; note that loading a word from the on-chip SRAM or off-chip DRAM costs much more energy (50x and 6400x, respectively) as compared with an ALU operation \cite{danowitz2012cpu, pedram2017dark}. Therefore, eliminating data movements/ communication significantly improves the overall performance.

An example of a program that could benefit from \gls{mvp} is illustrated in Fig. \ref{fig:cima}b. The program consists of multiple loops processing a dataset that is preloaded and mapped on \gls{mvp}. Each time a loop is called, the processor sends a (macro)-instruction to \gls{mvp}; the instruction is locally decoded and executed. The result is returned to the processor.   This feature occurs in multiple applications such as database management \cite{wu2005fastbit}, DNA sequencing \cite{soniefficient, cameron2014bitwise, lavenier2016dna}, and graph processing \cite{beamer2013direction}.

\subsection{Evaluation Results}

To evaluate \gls{mvp} architecture, its estimated performance is compared to a multicore architecture. The models and assumptions for the multicore architecture and \gls{mvp} are similar to those in \cite{hamdioui2015memristor, du2017implementation}; e.g., the multicore architecture consists of 4 cores (ALU only), two levels of caches (32 KB L1 and 256 KB L2) and 4 GB DRAM. The \gls{mvp} architecture consists of one core (ALU only), two levels of caches (32 KB L1 and 256 KB L2), 2 GB DRAM, and a \gls{mvp} with a 2 GB non-volatile crossbar memory with a modified read-out circuity (as explained in \cite{xie2017scouting}) in order to enable computation-in-memory.  Three metrics are used for the evaluation: (1) performance energy efficiency $\eta_{PE}$ (defined by MOPs/mW), (2) energy efficiency $\eta_{E}$ (defined by pJ/op), and (3) performance area efficiency $\eta_{PA}$ (defined by MOPs/mm$^2$).

Fig. \ref{fig:hpc} shows the results of the evaluation metrics for both architectures for different L1 and L2 cache misses (up to 60\%)and by assuming that 70\% of the program instructions can be accelerated on \gls{mvp} (\%Acc=0,7); i.e., the 30\% non-accelerated instructions is executed by the conventional processor and the 70\% accelerated part by \gls{mvp}; see Fig. \ref{fig:cima}. As \gls{mvp} architecture contains a conventional part (i.e., CPU, caches, DRAM and external memory), only 10x improvement is obtained with respect to the  performance-energy efficiency. MVP architecture also achieves one order of magnitude energy efficiency improvement in comparison with the multicore architecture, and has a higher performance area efficiency. Therefore, the MVP architecture has the potential of  realizing significant improvements, despite the high switching latency and low endurance of memristor devices. The improvements are the result of a significant reduction of cache and DRAM accesses, and the usage of non-volatile memory. The reduction of memory accesses leads to a lower latency and lower energy consumption, while the non-volatile memory reduces the static power practically to zero.

	\section{Memristive Devices for Automata Processing} 	\label{sec:MAP}
	Automata-based processing is widely used in diverse fields, including network security~\cite{Yu2006}, computational biology~\cite{Roy2016}, and data mining~\cite{Wang2016}. Its hardware implementation, referred to as \textit{automata processors (APs)}, has significant advantages over von Neumann architectures regarding throughput and energy efficiency as they enable computation-in-memory~\cite{Dlugosch2014,Bo2016,Subramaniyan2017a}. Memristive devices, which are the enablers of Resistive Random-Access Memories (RRAM) and computation-in-memory, are potential candidates for implementing the APs as it will be shown in this section. We will refer to this implementation as RRAM-AP. Moreover, it will be shown that RRAM-AP outperforms the two known hardware implementations of APs, being the Micron Automata Processor~\cite{Dlugosch2014} which is based on SDRAM, and the Cache Automation~\cite{Subramaniyan2017a} which is based on SRAM; we will refer to them by SDRAM-AP and SRAM-AP, respectively, to maintain the naming consistent with RRAM-AP. Next, we will first introduce basic knowledge and notations of automata. Subsequently, we propose a generic model for automata processors. Thereafter, we  present RRAM-AP implementation, and show its superiority.

	\subsection{Automata Basics}
	
	\begin{figure}[!t]
		\centering
		\subfloat[NFA]{
			\includegraphics[scale=1.2]{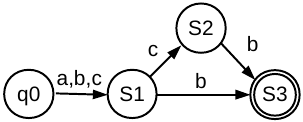}
			\label{fig:nfa}}
		\hfil
		\subfloat[Homogeneous automata]{
			\includegraphics[scale=1.2]{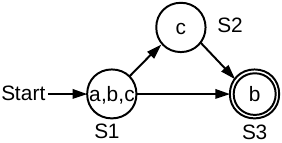}
			\label{fig:nfa:graph}}
		\caption{Example notations for NFAs and homogeneous automata.}
		\label{fig:graph}
	\end{figure}
	
	A \gls{nfa} can be represented by a 5-tuple: $(Q, \Sigma, \delta, q0, C)$. $Q$ represents a finite set of states (which are denoted with circles in the illustrative example of Fig.~\ref{fig:nfa}), $\Sigma$ is a finite set of possible input symbols (that can be used to generate an input sequence), $\delta$ is the transition function describing the set of possible transitions among the states, $q0$ is one of the states from $Q$ and presents the \textit{start state}, $C$ is a subset of $Q$ and contains the \textit{final states} or \textit{accepting states}; they are denoted with a double circle in the state diagram af Fig.~\ref{fig:nfa} as shown for the final state S3. 
	
	During operation (i.e., execution of an input sequence), some states can be \textit{active}; they are denoted by $P$. Initially, $P$ equals to $q0$. 
	At each processing step, the \gls{nfa} consumes one symbol $I$ from the input sequence. Based on $I$ and $\delta$, $P$ is updated. Once all symbols of the input sequence are processed, the NFA output is determined by $P$ and $C$.
	If $P\cap C\neq \emptyset$, then we say that the NFA \textit{accepts} the input sequence; otherwise, the sequence is \textit{rejected}. The acceptance of the input sequence can be represented by a Boolean value $A$.

	\textit{Homogeneous automaton} is a special type of NFA that is relatively easy to implement by APs~\cite{Dlugosch2014}. 
	It requires that a state can only be reached by transitions with the {\em same} input symbol(s). These input symbols belong to the \textit{symbol class} of this state. For example, in the NFA shown in Fig.~\ref{fig:nfa:graph}, S3 can be reached by two transitions (from S1 and S2, respectively) both with the same symbol $b$; $b$ belongs to the symbol class of S3.
	Here, the NFA shown in Fig.~\ref{fig:nfa} is a homogeneous automaton and can be therefore redrawn as depicted in Fig.~\ref{fig:nfa:graph}. Note that the input symbols are only related to the {\em states} in homogeneous automata and not the {\em state transitions} as is the case for normal NFAs; e.g., the symbol \textit{b} is not on the incoming edges/transition of the state S3 (see Fig.~\ref{fig:nfa}) but rather within the node representing S3 (see Fig.~\ref{fig:nfa:graph}).
	Any \gls{nfa} can be translated into its equivalent homogeneous automaton and therefore implemented using APs~\cite{Dlugosch2014}.

	

	\subsection{Generic Automata Processor Model}
	\begin{figure}[!t]
		\centering
		\includegraphics[width=\columnwidth]{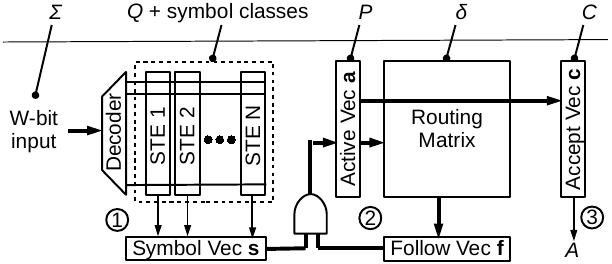}
		\caption{General architecture for automata processors.}
		\label{fig:arch}
	\end{figure}

Before implementing RRAM-AP, we need to understand the key operations conducted by an AP. Therefore, we next present a generic model for APs to identify these operations. This generic model is shown in Fig.~\ref{fig:arch} and consists of three major processing steps:
	
\begin{enumerate}
\item Input symbol processing: It decodes each symbol $I$ (presented with $W$ bits) of the input sequence by activating only one of the $2^W$ wordlines, 
and identifies all states that have an incoming transition occurring on $I$. These states and the remaining sates are presented by column vectors called \glspl{ste}, and are pre-configured based on $Q$ and the corresponding symbols (symbol class). Each STE presents one state of the $N$ states of $Q$. The result of this step is mapped to a vector called Symbol Vector $\mathbf{s}$.


\item Active state processing: It generates: (1) all the possible states that can be reached from the current active states $P$ (stored in a vector called Active Vector $\mathbf{a}$) based on these states and the transition function $\delta$ (stored in the routing matrix), and stores the result in the Follow Vector $\mathbf{f}$; (2) the next active states (i.e., Active Vector) by bit-wise ANDing $\mathbf{s}$ and $\mathbf{f}$.

		
\item Output identification: In order to decide about the value of $A$ (i.e., whether the input sequence is accepted or not), the intersection of $\mathbf{a}$ and the \textit{Accept Vector} $\mathbf{c}$ (pre-configured based on $C$) is checked. That is, if $P\cap C\neq \emptyset$, then $A=1$ (accept), otherwise $A=0$ (reject).
\end{enumerate}
	
Next we will elaborate the above three processing steps. 

\subsubsection{Input symbol processing}
As mentioned, the purpose of this is to calculate the Symbol Vector $\mathbf{s}$ for each input symbol. This is done based on the selected row (from the $2^W$ rows) and the configuration of STEs. Let's assume that for each input symbol, an \textit{Input Vector} $\mathbf{i}$ of $2^W$ elements is generated where only one element is high (corresponding to the selected wordline); the remaining elements are 0. In addition, assume that the configuration of STEs can be presented by a matrix $\mathbf{V}$ where each column $\mathbf{V_n}$ presents the STE of the state $n$. Then the $n$th element of the Symbol Vector $\mathbf{s}$ corresponding to $\mathbf{V_n}$ can be calculated as:

\begin{equation}
	s[n] = \mathbf{i} \cdot \mathbf{V_n} = \sum_{k=0}^{2^W}i[k]v_n[k],\ \forall n \in [1, N]
	\label{eq:ste}
\end{equation}

In this equation, the addition and the multiplication represent the Logic OR and AND, respectively. For the example of Fig.~\ref{fig:nfa:graph}, if we assume $\Sigma=\{a,b,c,d\}$, then, 
	$$\mathbf{V}=\begin{bmatrix}\mathbf{V_1} & \mathbf{V_2} & \mathbf{V_3}\end{bmatrix} = \begin{bmatrix} 1 & 0 & 0 \\ 1 & 0 & 1 \\ 1 & 1 & 0 \\ 0 & 0 & 0\end{bmatrix}.$$
This means that S1's symbol class is $\{a,b,c\}$, S2's is $\{b\}$, and S3's is $\{c\}$. If we further assume that the current input symbol is $b$, then $\mathbf{i}=[0\ 1\ 0\ 0]$, and $\mathbf{s}=[1\ 0\ 1]$. This means that $b$ is in the symbol classes of S1 and S3.

	\subsubsection{Active states processing}
This step calculates the Follow Vector $\mathbf{f}$ which presents the possible states that can be reached from the current active states stored in the Active Vector $\mathbf{a}$. The transition function is implemented by the routing matrix as shown in Fig.~\ref{fig:arch}, and can be conceptually presented as a two-dimensional vector $\mathbf{R}$. Hence, the $n$th element of Follow Vector $\mathbf{f}$ can be calculated as:
	
\begin{equation}
f[n] = \mathbf{a} \cdot \mathbf{R_n} = \sum_{i=0}^{N-1}a[i]R_n[i],\ \forall n \in [1, N].
\label{eq:follow}
\end{equation}

The interpretation of the addition and the multiplication in this equation is the same as in Equation (1). The next active states (to be also stored in the Active Vector $\mathbf{a}$) are easily calculated by using bitwise AND operation.

\begin{equation}
a[n] = f[n]\ \&\ s[n], \ \forall n \in [1, N].
\label{eq:active}
\end{equation}

For the example of Fig.~\ref{fig:nfa:graph}, the matrix $\mathbf{R}$ that belongs to the transit function is
$$\mathbf{R}=\begin{bmatrix}\mathbf{R_1} & \mathbf{R_2} & \mathbf{R_3}\end{bmatrix} = \begin{bmatrix}0 & 1 & 1 \\ 0 & 0 & 1 \\ 0 & 0 & 0\end{bmatrix}.$$ 
This means that S1 cannot be reached from all the states ($\mathbf{R_1}$), S2 can only be reached from S1 ($\mathbf{R_2}$), and S3 from both S1 and S2 ($\mathbf{R_3}$). For $\mathbf{a}=[1\ 0\ 0]$ (only S1 is active), $\mathbf{f}=[0\ 1\ 1]$ according to Equation (2). This means S2 and S3 are reachable states from the active states. If we assume the next input symbol is $b$, which leads to $\mathbf{s}=[1\ 0\ 1]$ as discussed above, then the new active vector $\mathbf{a}=[0\ 0\ 1]$ according to Equation (3). This means that S3 becomes the next active state.		
		
\subsubsection{Output identification}
The output value $A$ of NFA  is easily calculated using the Active Vector $\mathbf{a}$ and the Accept Vector $\mathbf{c}$. The former stores the active states generated by the input sequence while the later stores the defined accepting states of NFA. 
\begin{equation}
A = \mathbf{a} \cdot \mathbf{c}^\top = \sum_{n=0}^{N-1}a[n]c[n].
\label{eq:acc}
\end{equation}

$A=1$ means that the input symbol sequence is accepted by the \gls{nfa}; otherwise, the string is rejected. For the example of Fig.~\ref{fig:nfa:graph}, $\mathbf{c}=[0\ 0\ 1]$. This means only S3 is an accepting state. If we assume the same example as above ($\mathbf{a}=[0\ 0\ 1]$), then $A=1$.
	
\subsection{RRAM-AP Implementation}
The automata processing model described above contains only two types of logic operations, which are vector dot product (Equation \ref{eq:ste}, \ref{eq:follow}, and \ref{eq:acc}) and vector bit-wise AND (Equation~\ref{eq:active}). In practice, we cannot implement the complete routing matrix of Equation~\ref{eq:follow}, as it requires too much resource. SDRAM-AP and SRAM-AP both use hierarchical routers to implement the routing matrix. Their implementations do not support all NFA transitions; nevertheless, there is enough flexibility to route all possible transitions of typical applications~\cite{Dlugosch2014,Subramaniyan2017a}. While SDRAM-AP does not reveal many implementation details, SRAM-AP uses a two-level structure that consists of global and local switches~\cite{Subramaniyan2017a}. These global and local switches also conduct vector dot product operations.
	
\begin{figure}[!t]
	\centering
	\subfloat[Used as STEs]{
		\includegraphics[scale=2]{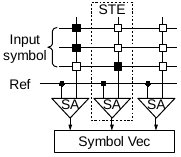}%
		\label{fig:switch:ste}}
	\hfil
	\subfloat[Used as routers]{
		\includegraphics[scale=2]{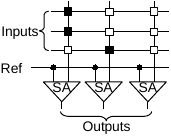}%
		\label{fig:switch:router}}
	\caption{Vector dot product operator used as switches and STEs.}
	\label{fig:switch}
\end{figure}

For our implementation, we adopt SRAM-AP's for the routing matrix, use the hardware structure shown in Fig.~\ref{fig:switch:ste} for STEs, and the one in Fig.~\ref{fig:switch:router} both for global and local switches. The black and white boxes represent different configuration bits. Each column generates the vector dot product of the input vector and the configuration bits of this column.
	
	\begin{figure}[!t]
		\centering
		\subfloat[Program circuit]{
			\includegraphics[scale=2]{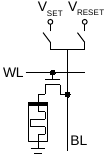}
			\label{fig:program}}
		\hfil
		\subfloat[RRAM cell]{%
			\includegraphics[scale=2]{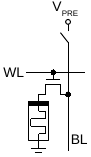}%
			\label{fig:switch:rram}}
		\hfil
		\subfloat[SRAM cell~\cite{Subramaniyan2017a}]{
			\includegraphics[scale=2]{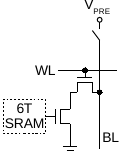}%
			\label{fig:switch:sram}}
		\caption{Different implementations of a configurable bit.}
		\label{fig:bit}
	\end{figure}
	
An NFA is configured to RRAM-AP by programming RRAM devices to either low or high resistance. We use one transistor and one RRAM device (1T1R) to implement a configurable bit as shown in Fig.~\ref{fig:switch:rram}. During the configuration, the word line WL selects the row to be programmed, and the programming voltage is applied to the bit line BL as shown in Fig.~\ref{fig:program}. The programming voltage can be either SET or RESET voltage. Logic 1 corresponds to the memristor's low resistance, and logic 0 to high resistance. The bit line is pre-charged before evaluation, and the word lines are selected, e.g., by the input symbols. Note that for the routing matrix, multiple word lines can be activated in parallel. The vector dot product is calculated when all the word lines are set; if all the corresponding selected cells contain a high resistance (i.e., logic 0), then the pre-charged bit line remains high, and the sense amplifier (SA) will read a logic 0 (inverted output). Similarly, if at least one of the cells contains a low resistance (i.e., logic 1), then BL will be discharged. The SA's output will subsequently be a logic 1.
	
The characteristics of memristors provide opportunities for RRAM-AP to outperform previous designs. For example, SRAM-AP uses eight transistors to implement the configurable bit as shown in Fig.~\ref{fig:switch:sram} \cite{Subramaniyan2017a}, whose area is much larger than the 1T1R structure. In addition, the SRAM cells also suffers from leakage power. As memristors are non-volatile devices, RRAM-AP can resume the last configured NFA after shut down and reboot without reprogramming it. On the other hand, RRAM-AP also inherits some drawbacks, such as the longer and power-hungry programming phase, and lower endurance, in comparison with SDRAM and SRAM.
	
\subsection{Preliminary Results}
	
The APs can be built by using only vector dot product and bit-wise AND operators. Except for the vector dot product operator, we assume that the remaining part of RRAM-AP is implemented in a similar way as SRAM-AP (incl. bit-wise AND, wiring, and sense amplifiers). Hence, we compare only the dot product operator. Note that SRAM-AP outperforms SDRAM-AP regarding the throughput and energy consumption; therefore, we limit our comparison to SRAM-AP.
	
	\begin{figure}[!t]
		\centering
		\subfloat[Simulated circuit]{
			\includegraphics[scale=2.2]{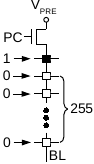}
			\label{fig:circuit}}
		\hfil
		\subfloat[SPICE simulation result]{
			\includegraphics[width=2.4in]{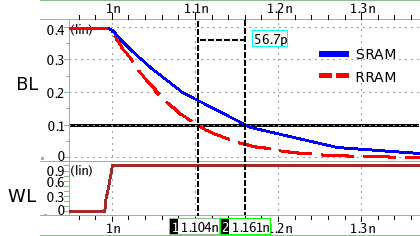}
			\label{fig:spice}}
		\caption{SPICE simulation results of a vector dot product operator.}
		\label{fig:discharge}
	\end{figure}
	
The simulated circuit consists of a single vector dot product operator with a length of 256 as shown in Fig.~\ref{fig:circuit}. We use \SI{32}{\nano\meter} PTM model for CMOS transistors and ASU model~\cite{Chen2015b} for RRAM. We configure RRAM's parameters based on a two-state device, similarly as presented in~\cite{Tran2011}, e.g., the RRAM's high and low resistances are approximately \SI{100}{\mega\ohm} and \SI{1}{\kilo\ohm} respectively; the SET and RESET threshold voltages are \SI{1.3}{\volt} and \SI{0.5}{\volt}. To simulate the slowest discharge process, only the first cell is configured to logic 1 (indicated by the black box), and the remaining 255 cells are configured to be 0 (indicated by white boxes).
The bit line BL is pre-chared to \SI{0.4}{\volt} (lower than RRAM's threshold voltages). When BL is discharged to \SI{0.1}{\volt}, the sense amplifier (not included in the circuit) will read a 1. The reference voltage of the SA is set to \SI{0.25}{\volt}.
	
The HSPICE simulation results are shown in Fig.~\ref{fig:spice}. 
The word line WL is enabled at \SI{1}{\nano\second}, and then BL starts discharging. BL's voltages in SRAM and RRAM-based designs are illustrated with solid blue line and dashed red line, respectively.
The discharge time through RRAM (\SI{104}{\pico\second}) is 35\% less than the SRAM-based implementation (\SI{161}{\pico\second}). This is mainly because transistors have relatively large intrinsic capacitance. During bit-line discharge, the RRAM cell of Fig.~\ref{fig:switch:rram} has only one transistor in its path while the SRAM-based design has two (See Fig.~\ref{fig:switch:sram}). The energy consumed during the charge and discharge processes is \SI{2.09}{\femto\joule} for the RRAM-based design and \SI{5.16}{\femto\joule} for the SRAM-based design. The former is 59\% less than the latter. Considering that the remainder part of RRAM-AP is implemented in a similar way as SRAM-AP, RRAM-AP outperforms SRAM-AP at the chip level regarding latency, energy, and area.

\section{Conclusion}
\label{sec:conclusion}
In this work, we have discussed two potential applications of memristive devices and computation-in-memory, i.e., Memristive Vector Processor and RRAM Automata Processor. Memristors' unique properties provide us an important opportunity to improve conventional designs at both architectural and device level. However, the drawbacks of memristor technology, such as the impact of endurance, require further research.


	\section*{Acknowledgment}
	This work was supported by the European Union’s Horizon 2020 Research and Innovation Program through the project MNEMOSENE (Grant 780215).
	
	
	
	%

\end{document}